\documentclass[11pt]{article}
\usepackage{times}
\usepackage{graphics}
\usepackage{color}
\usepackage{amssymb}
\usepackage{amsmath}
\usepackage{ifthen}
\usepackage{calc}

\newtheorem{theorem}{Theorem}

\newcommand{\TMSF}[1]{\ensuremath{T_{\mathrm{MSF}}\ifthenelse{\equal{#1}{}}{}{(#1)}}}
\newcommand{\TCoarse}[1]{\ensuremath{T_{\mathrm{Coarse}}\ifthenelse{\equal{#1}{}}{}{(#1)}}}
\newcommand{\TLabel}[1]{\ensuremath{T_{\mathrm{Label}}\ifthenelse{\equal{#1}{}}{}{(#1)}}}

\begin{document}

\begin{center}

{\Large\textbf{Optimal Component Labeling Algorithms\\[\medskipamount]for Mesh-Connected Computers and VLSI}}
\bigskip

{\large Quentin F. Stout}
\bigskip

Computer Science and Engineering\\
University of Michigan
\end{center}

\vspace{0.3in}

\centerline{\textbf{Abstract}}
\medskip

\noindent
Given an undirected graph $G$ of $n$ weighted edges, stored one edge per processor in a square mesh of $n$ processors, we show how to determine the connected components and a minimal spanning forest in $\Theta(\sqrt{n})$ time.
More generally, we show how to solve these problems in $\Theta(n^{1/d})$ time when the mesh is a $d$-dimensional cube, where the implied constants depend upon $d$.
\bigskip


\section{Introduction}

Note: these results were first announced by the author in 1984~\cite{1984}, and were obtained contemporaneously by John Reif.
We intended to publish a joint paper, but never got around to doing so.
Since the results have been utilized many times, and over the years I've explained the algorithm to several people, I thought it useful to make this note available.
I kept the title of the original announcement despite the fact that very few talk about algorithms for VLSI anymore.
\bigskip

\begin{figure}
\begin{center}
\begin{minipage}[b]{2.1in}
\centerline{\resizebox{2in}{!}{\includegraphics{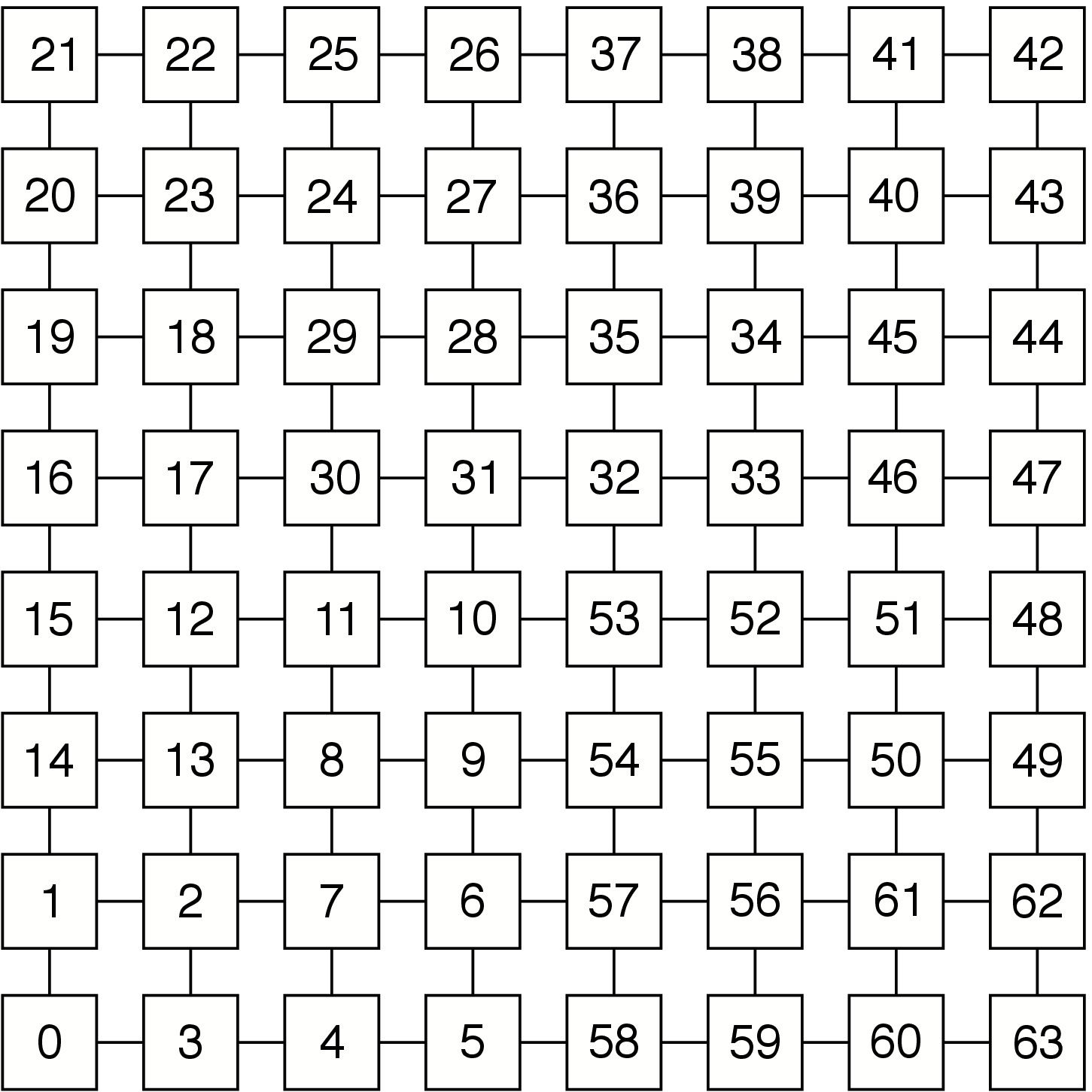}}}
\centerline{a)}
\end{minipage}
\hspace{0.5in}
\begin{minipage}[b]{2.6in}
\centerline{\resizebox{2.5in}{!}{\includegraphics{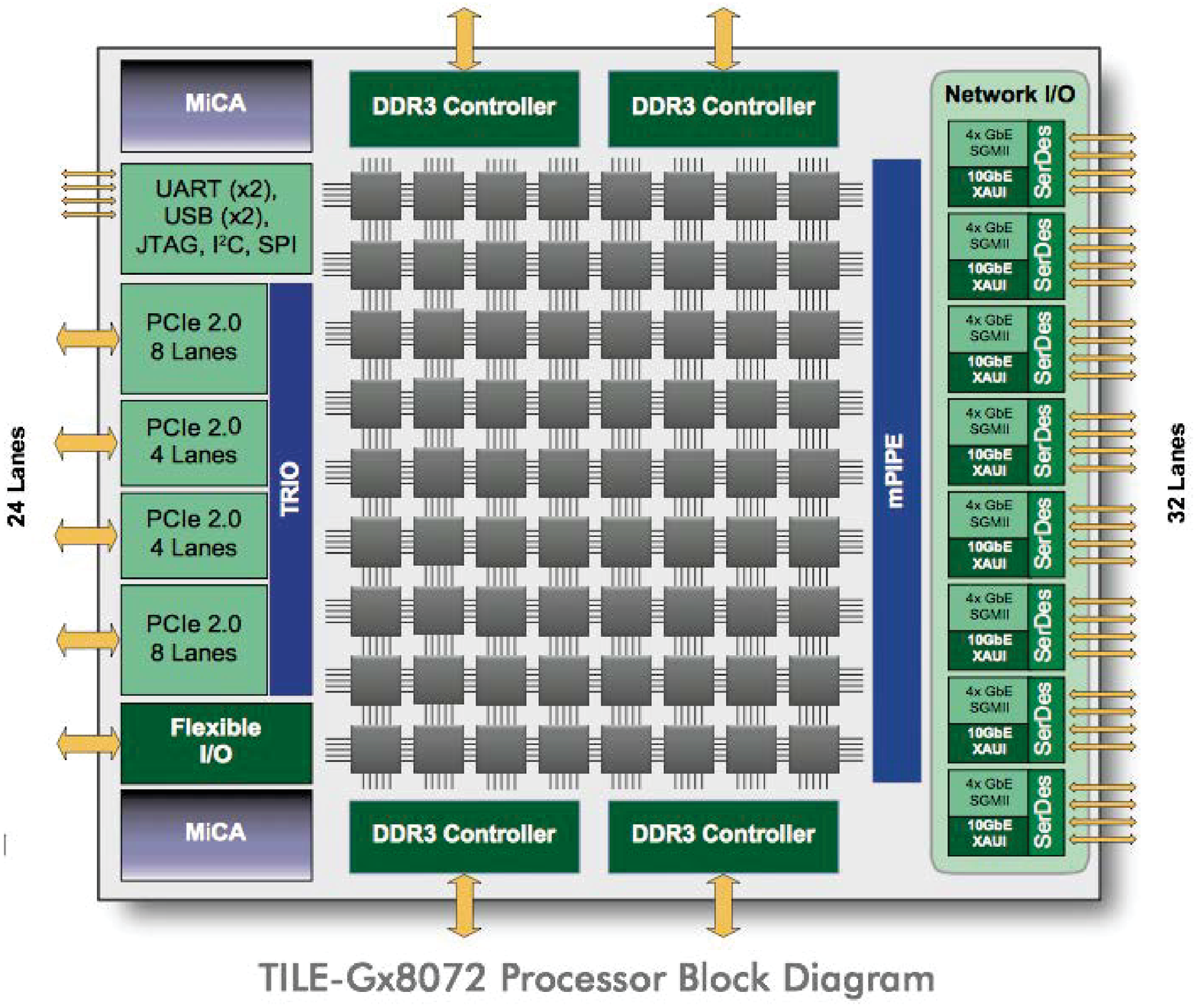}}}
\centerline{b) Tilera processor}
\end{minipage}
\end{center}

\caption{Abstract and Implemented Meshes} \label{fig:mesh}
\hrulefill
\end{figure}

We give algorithms for determining a minimal spanning forest, and the connected components, of an undirected graph stored on a mesh-connected computer.
Figure~\ref{fig:mesh} a) shows a 2-dimensional mesh-connected computer.
Each processor can directly communicate with its 4 neighbors in unit time, and messages to further away processors must be passed from neighbor to neighbor.
A fine-grained model is used, where each processor has a fixed number of words of memory and does fundamental operations in unit time.
This is a slight extension of cellular automata, where each automaton has only a fixed number of bits of memory.

Meshes have long been used as a fundamental form of parallel computer, and were especially attractive for VLSI design because very little area was wasted on connections~\cite{Ullman84}.
Meshes, though not as fine-grained, are once again important for chip design.
Recent interest is motivated by the fact that the distance signals travel determines the time and energy needed.
Energy consumption is becoming a dominant constraint of chip design, and chips with large numbers of simple processors connected as a mesh are becoming available~\cite{Tilera} (see Figure~\ref{fig:mesh} b)).
Several proposed designs of processing chips for exascale computers assume there will be many RISC cores connected as a 2-d mesh~\cite{DOE:SciGrandChal,DOE:ExaProgChal}.

The input for our problems is the edges of an undirected weighted graph $G=(V,E)$ with $n$ edges, stored one per processor on a $\sqrt{n} \times \sqrt{n}$ mesh.
Each edge $(u,v)$ has an associated weight $w(u,v) \geq 0$, where for an unweighted graph $w(u,v)=1$ is implied.
Whenever edges are being moved their weight goes with them.
We assume there is an ordering on the labels of the vertices.
To simplify exposition, assume that each edge is represented twice, so that an edge between vertices $u$ and $v$ is stored as $(u,v)$ and as $(v,u)$.
Further, for every vertex $v$ there is a self-loop, i.e., an edge of the form $(v,v)$.
This guarantees that every vertex is represented.
We will show:

\begin{theorem}
Given an undirected weighted graph $G=(V,E)$ with $n$ edges, stored one edge per processor on an $\sqrt{n} \times \sqrt{n}$ mesh computer, in $\Theta(\sqrt{n})$ time one can find a minimal spanning forest and label the connected components.
\end{theorem}

\begin{figure}

\centerline{\resizebox{5in}{!}{\includegraphics{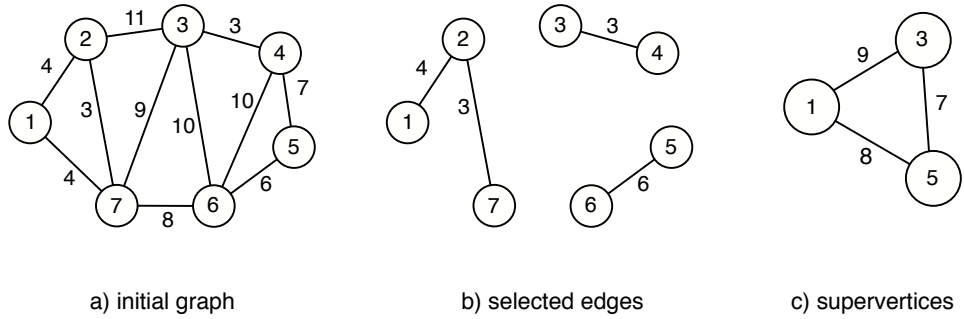}}}

\caption{Coarsening} \label{fig:coarsen}

\hrulefill
\end{figure}

The algorithm to find a minimal spanning forest (MSF) uses a recursive approach originally used by Bor\.{u}kva and which has been rediscovered by many others (including Prim).
For each vertex an edge is selected, forming a forest but not necessarily a spanning forest.
The edges selected become part of the MSF, and the trees are supervertices used as the vertices in the next iteration.
For example, for the graph in Figure~\ref{fig:coarsen} a), the edges selected are shown in b).
This is known as \textit{coarsening}.
The edge between supervertices $U$ and $V$ is one having minimal
weight among the edges connecting a vertex in $U$ with one in $V$.
Ties can be broken arbitrarily.
The result is shown in c).
If a supervertex is not connected to any other others then it is finished.
After a coarsening step the number of unfinished supervertices is at most 1/2 the number of vertices since each such supervertex contains at least two vertices.

Throughout, all sorting is into a space-filling curve ordering, such as the Hilbert ordering illustrated in Figure~\ref{fig:mesh} a).
Sorting in a $\sqrt{n} \times \sqrt{n}$ mesh can be done in $\Theta(\sqrt{n})$ time~\cite{ThKu77}.
\bigskip

\noindent
\textbf{Minimal Spanning Forest Algorithm:} \nopagebreak
\begin{list}{\arabic{enumi})}{\usecounter{enumi}}
\item Do coarsening 5 times, leaving at most $n/32$ remaining supervertices.

\item In each quadrant of the mesh, recursively solve the MSF problem for the supervertices, using only the edges in the quadrant.
The number of edges in a quadrant's MSF is no more than the number of supervertices, so for all of the quadrants combined the number of edges is at most
$4\cdot(n/32) = n/8$.

\item Move these edges to a submesh of size $n/8$ and recursively solve the MSF problem in this submesh.  This uses the fact that a MSF of the union of the MSFs of the subgraphs is a MSF of the entire graph.

\end{list}
\noindent
Step 1) reduces the number of vertices, then step 2) reduces the number of edges. 
Without 2), even though the graph has fewer vertices at the end of 1), the number of edges may not have been reduced much, so step 3) would not have been in a small submesh.
\bigskip

\noindent \textbf{Coarsening:}
\begin{list}{\alph{enumi})}{\usecounter{enumi}}
\item For every vertex $u$ select the edge $(u,v)$ of minimal weight among all incident edges, where if there are ties select the one where $v$ has minimal label.
This is an edge in the MSF.
If a vertex has no edges then it is removed from further consideration and already has its label.

\item The selection rule insures that no cycles are created and hence the edges selected form a forest.
Label the trees, creating the supervertices.

\item For every pair of supervertices $U$, $V$, choose one of the edges of minimal weight among those with one endpoint in $U$ and one in $V$. 
Ties can be broken arbitrarily.
This is the edge between $U$ and $V$ in the coarsened graph.
See Figure~\ref{fig:coarsen} c).
\end{list}

\noindent
While the trees created during coarsening are undirected, labeling them involves creating directed subtrees, with each edge pointing towards the root of its subtree.
Furthermore, weights are ignored.
\bigskip

\begin{figure}
\centerline{\resizebox{4.3in}{!}{\includegraphics{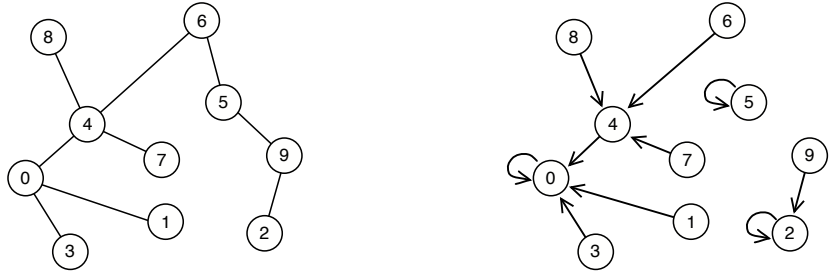}}}
\caption{Forming Directed Subtrees} \label{fig:subtrees}
\hrulefill
\end{figure}

\noindent
\textbf{Labeling Trees:} \nopagebreak
\begin{list}{\roman{enumi})}{\usecounter{enumi}}

\item Using only the tree edges selected in coarsening, for each vertex $u$ determine the neighboring vertex $v$ with the smallest label and create the directed edge $(u,v)$, where the edge is $(u,u)$ if $u$'s label is less than all of its neighbors'.
This directed edge represents $u$ in the following steps.
The edges selected from each tree form directed subtrees, as shown in Figure~\ref{fig:subtrees}.

\item Sort the directed edges by the label of the vertex being pointed at.
Because sorting is in space-filling curve order, vertices in any quadrant can only point to vertices in the same quadrant or a quadrant of smaller index.

\item For every vertex, find the label of the root of its subtree.
This is done in a bottom-up fashion, using $2\times2$ meshes, then $2^2\times 2^2$ meshes, etc.
For vertex $u$ let $M(u,i)$ be the $2^i \times 2^i$ mesh containing $i$, and let $A(u,i)$ be its greatest ancestor (node closest to the root on the path from $u$ to the root) represented in $M(u,i)$.
Thus if $x$ is $A(u,i)$'s parent then edge $(A(u,i),x)$ is in $M(u,i)$ while $x$'s representative is not in $M(u,i)$.
Initially $A(u,0)=v$, where $v$ is the neighbor selected in step i).
To determine $A(u,i\!+\!1)$, note that if the parent of $A(u,i)$ is in $M(u,i\!+\!1)$ then it must be in a quadrant of smaller index since its label is smaller than $A(u,i)$'s.
Thus by 3 iterations of every quadrant of $M(u,i\!+\!1)$ searching in the quadrant preceding it,
$A(u,i\!+\!1)$ can be determined.
This is not a recursive call, merely a merging operation, thus this step can be completed in $\Theta(\sqrt{n})$ total time.

\item In the preceding step we don't actually keep track of $i$, but rather just keep updating a value $A(u)$.
At the end, for each vertex $u$ for which $A(u)=u$ and no neighbors point to $u$, if $u$ has neighbors in the graph (such as vertex 5 in the example), choose an arbitrary neighbor $v$ and set $A(u)=A(v)$
If it has no neighbors then $u$'s label is $u$ and it is finished.
The number of unfinished subtrees is at most 1/2 the number of vertices since each unfinished subtree represents at least 2 vertices.

\item The subtrees are now supervertices for this tree labeling routine, and for any pair of supervertices either they are not adjacent in the original tree or there is a unique edge in the original tree that connects them.
Move the the edges connecting supervertices to a submesh and recursively call the routine starting at step ii).
\end{list}

To analyze the time let \TMSF{n}, \TCoarse{n}, \TLabel{n} denote the time for finding the minimal spanning forest, coarsening, and labeling the trees, respectively, on an $\sqrt{n} \times \sqrt{n}$ mesh.
\begin{eqnarray*}
  \TLabel{n} & = & \TLabel{n/2} + \Theta(\sqrt{n})\\
          & = & \Theta(\sqrt{n})\\[\bigskipamount]
  \TCoarse{n} & = & \TLabel{n} + \Theta(\sqrt{n})\\
          & = & \Theta(\sqrt{n})\\[\bigskipamount]
 \TMSF{n} & = & 5 \TCoarse{n} + \TMSF{n/4} + \TMSF{n/8} + \Theta(\sqrt{n})\\
          & = & \Theta(\sqrt{n})
\end{eqnarray*}

For component labeling, at the end of the MSF algorithm one can do tree labeling on the edges selected to determine the component labels. 

Finally, all of the above can be extended to $d$-dimensional meshes in a straightforward manner, the only difference being that $2d+1$ vertex reductions are needed in step 1). 
This gives
\begin{theorem}
Given an undirected weighted graph $G=(V,E)$ with $n$ edges, stored one edge per processor on a cubical $d$-dimensional mesh computer, in $\Theta(n^{1/d})$ time one can find a minimal spanning forest and label the connected components, where the implied constants depend upon $d$.
$\Box$
\end{theorem}


\begin{thebibliography}{9}

\bibitem{DOE:SciGrandChal}
DOE,
``Scientific grand challenges: architectures and technology for extreme scale computing'',
\textit{Report of DOE Workshop}, Dec.\ 2009

\bibitem{DOE:ExaProgChal}
DOE,
``Exascale programming challenges'',
\textit{Report of DOE Workshop}, July 2011.


\bibitem{1984} Stout, Q.F.,
 ``Optimal component labeling algorithms for mesh-connected computers and VLSI'',
 \textit{Abstracts AMS} 5, 1984, 148.
 
 \bibitem{ThKu77} Thompson, C.D. and Kung, H.T.,
 ``Sorting on a mesh-connected parallel computer'', 
 \textit{Comm.\ ACM} 20, 1977, 263--271.

 
 \bibitem{Tilera} Tilera Corporation, 
 ``TILE-Gx8072 Processor Product Brief'',\\
 www.tilera.com/sites/default/files/productbriefs/TILE-Gx8072\_PB041-04\_WEB.pdf

\bibitem{Ullman84}  Ullman, J.D.,
 \textit{Computational Aspects of VLSI}, 1984,
 Computer Science Press, Maryland.

\end{thebibliography}
\end{document}